\documentclass[preprint,12pt,a4paper]{elsarticle}

\usepackage{multirow}

\usepackage{epsfig}

\usepackage{amssymb}

\usepackage{float}

\usepackage{booktabs} 

\usepackage{multicol}

\biboptions{compress}

\begin{document}

\begin{frontmatter}

\title{Soft Pomerons and the Forward LHC Data}

\author[label1]{M. Broilo}
\author[label1]{E.G.S. Luna}
\author[label2]{M.J. Menon}

\address[label1]{Instituto de F\'{\i}sica, Universidade Federal do Rio Grande do Sul \\
Caixa Postal 15051, 91501-970, Porto Alegre, RS, Brazil}
\address[label2]{Instituto de F\'{\i}sica Gleb Wataghin, Universidade Estadual de Campinas - UNICAMP \\
13083-859 Campinas, SP, Brazil}


\begin{abstract}
Recent data from LHC13 by the TOTEM Collaboration on $\sigma_{tot}$ and $\rho$
have indicated disagreement with all the Pomeron model predictions by
the COMPETE Collaboration (2002). On the other hand, as recently demonstrated by Martynov and Nicolescu
(MN), the new $\sigma_{tot}$ datum and the unexpected decrease in the $\rho$ value
are well described by the maximal Odderon dominance at the highest energies. 
Here, we discuss the applicability of Pomeron dominance through fits to the \textit{most complete
set} of forward data from $pp$ and $\bar{p}p$ scattering.
We consider an analytic parametrization for $\sigma_{tot}(s)$ consisting of non-degenerated
Regge trajectories for even and odd amplitudes (as in the MN analysis) and two Pomeron
components associated with double and triple poles in the complex angular momentum plane.
The $\rho$ parameter is analytically determined by means of dispersion relations.
We carry out fits to $pp$ and $\bar{p}p$ data on $\sigma_{tot}$ and $\rho$
in the interval 5 GeV -- 13 TeV (as in the MN analysis). Two novel aspects of our analysis are:
(1) the dataset comprises all the accelerator data below
7 TeV and we consider \textit{three independent ensembles} by adding: either only the TOTEM data (as in the MN analysis), or only the ATLAS data, or both sets; (2) 
in the data reductions to each ensemble, uncertainty regions are evaluated through
error propagation from the fit parameters, with 90 \% CL.
We argument that, within the uncertainties, this analytic model corresponding to soft Pomeron dominance, 
does not seem to be
excluded by the \textit{complete} set of experimental data presently available.
\end{abstract}


\end{frontmatter}

\begin{multicols}{2}

\section{Introduction}

The forward elastic hadron scattering is characterized by two quantities,
the total cross section and the $\rho$ parameter, which can be expressed, at high energies, in terms
of the amplitude $\mathcal{A}$ by \cite{pred}
\begin{eqnarray}
\sigma_{\mathrm{tot}}(s) = \frac{\mathrm{Im}\,\mathcal{A}(s,t=0)}{s},
\label{st}
\end{eqnarray}
\begin{eqnarray}
\rho(s) = \frac{\mathrm{Re}\,\mathcal{A}(s,t=0)}{\mathrm{Im}\,\mathcal{A}(s,t=0)},
\label{rho}
\end{eqnarray}
where $s$ and $t$ are the energy and momentum transfer squared in the center of mass system,
respectively.

In the Regge-Gribov formalism \cite{collins,fr,land}, the singularities in the complex angular momentum
$J$-plane ($t$-channel) are associated with the asymptotic behavior of the elastic scattering
amplitude in terms of the energy ($s$-channel). 
In the general case, associated with a pole of order $N$, the contribution to the 
imaginary part of the \textit{forward} amplitude
in the $s$-channel
is $s^{\alpha_0} \ln^{N-1}(s)$, where $\alpha_0$ is the intercept of the trajectory
(see Appendix B in \cite{fms17b} for a recent short review).
Therefore, for the total cross section we have 
\begin{eqnarray}
\sigma_{\mathrm{tot}}(s)  \propto s^{\alpha_0 - 1}\ln^{N-1} s,
\nonumber
\end{eqnarray}
and the following possibilities connecting the singularities at $J = \alpha_0$ and the asymptotic behavior:
simple pole ($N=1$)  $\Rightarrow$\ \ $\sigma \propto s^{\alpha_0-1}$;
double pole ($N=2$) at $\alpha_0=1$\ \ $\Rightarrow$\ \ $\sigma \propto \ln(s)$;
triple pole ($N=3$) at $\alpha_0=1$\ \ $\Rightarrow$\ \ $\sigma \propto \ln^2(s)$.

Most Pomeron models (even under crossing) consider leading contributions associated
with either a simple pole at $J=\alpha_0$ (for example, Donnachie and Landshoff \cite{dl13}) or
a triple pole at $J=1$ (as selected by the COMPETE Collaboration \cite{compete1,compete2}
and used in successive editions of the Review of Particle Physics \cite{pdg16}).

Recently, new experimental information on $\sigma_{\mathrm{tot}}$ and $\rho$ from LHC13 
were presented by the TOTEM Collaboration \cite{totem1,totem2}:
\begin{eqnarray}
\sigma_{\mathrm{tot}} &=& 110.6 \pm 3.4\ \mathrm{mb}, \nonumber \\
\rho &=& 0.10 \pm 0.01\ \mathrm{and}\ 0.09 \pm 0.01.
\label{totem13}
\end{eqnarray}

Remarkably, these two results seem not to be simultaneously described
by conventional models based on Pomeron exchanges,
as all those included in the detailed analysis by the COMPETE
Collaboration in 2002 (see Figure 18 in \cite{totem2}). 
On the other hand,
the odd-under-crossing asymptotic contribution,
introduced by Lukazsuk and Nicolescu \cite{odderon} and
named Odderon \cite{oddname}, provide quite good descriptions of the
experimental data, as predicted by the Avila-Gauron-Nicolescu model \cite{agn} and
demonstrated very recently in the forward analysis by Martynov and Nicolescu (MN)
on $pp$ and $\bar{p}p$ scattering in the interval 5 GeV -- 13 TeV
\cite{martynico}.

However, in their data reductions, MN consider only the TOTEM data at the LHC energy region
(excluding the ATLAS data at 7 and 8 TeV \cite{atlas7,atlas8}) and although the resulting curves cross the central values of the data
at 13 TeV, there is no
reference to uncertainty regions in the theoretical results.

Now, given the tension between the TOTEM and ATLAS data at 7 TeV and mainly
8 TeV \cite{fms17a}, the strict exclusion of the ATLAS data may not be a well justified procedure.
Moreover, since the uncertainties in the TOTEM data are essentially systematic
(and not statistical), the agreement between theoretical result and central value
may have a limited significance (see Appendix A in \cite{fms17b}).

Also very recently, the data at 13 TeV have been analyzed 
in the context of a two-component eikonal model by Khoze, Martin and 
Ryskin \cite{kmr1}, who also discuss inconsistencies
relating  maximal Odderon and the black disk limit \cite{kmr2}.

In the present work, our purpose is to discuss the applicability of a Pomeron dominance at the highest energies,
by taking into account: (1) all the experimental data presently available on $\sigma_{tot}$ and $\rho$
from $pp$ and $\bar{p}p$ in the interval 5 GeV -- 13 TeV; (2) the uncertainties involved in
the data reductions, interpolations and extrapolations. 

To this end, we consider a parametrization for $\sigma_{tot}(s)$
consisting of two simple poles Reggeons, even and odd ($a_2/f_2$ and $\rho/\omega$ mesonic trajectories, respectively)
and two Pomeron contributions, associated with double and triple poles in the $J$-plane, all the poles
corresponding, respectively, to powers (RR), logarithmic (L1) and logarithmic-squared (L2) dependences for the total cross section.
Inspired by the COMPETE notation we shall denote RRL1L2 model.

Following \cite{fms17a}, we consider three ensembles of
$pp$ and $\bar{p}p$ data above 5 GeV, all of them comprising the same dataset
in the region below 7 TeV, but distinguished by the addition of either the TOTEM
data, or ATLAS data, or both sets.

The main question to be discussed here can be put as follows: Did the forward LHC data exclude the Soft Pomeron?

Based on the data reductions, the fit uncertainty region with 90 \% CL, the uncertainties in 
the $\sigma_{tot}$ and $\rho$ data
at 13 TeV and further arguments, we are led to conclude
that the RRL1L2 model is not  
excluded by the complete set of experimental data presently available
on forward $pp$ and $\bar{p}p$ scattering above 5 GeV.

This paper is organized as follows. After introducing the analytic model in Sect. 2, we present
the fit procedures and results in Sect. 3, the discussions on all the results in Sect.
4 and our conclusions and final remarks in Sect. 5.

\section{Analytic Model - RRL1L2}

The analytic parameterization for the total cross section is given by 

\begin{eqnarray}
\sigma_{tot}(s) &=& a_1 \left[\frac{s}{s_0}\right]^{-b_1} + \tau a_2 \left[\frac{s}{s_0}\right]^{-b_2}
\nonumber \\
&+& A \ln\left(\frac{s}{s_0}\right) 
+ B \ln^2\left(\frac{s}{s_0}\right),
\label{stmodel}
\end{eqnarray}
where $a_1$, $b_1$, $a_2$, $b_2$, $A$ and $B$ are free fit parameters, $\tau = -1$ for $pp$, $\tau = +1$ for $\bar{p}p$ and $s_0$ is an energy scale. Here, as in the recent analyses by Fagundes, Menon and Silva \cite{fms17b,fms17a}, we assume this scale
as fixed at the physical threshold for scattering states,
\begin{equation}
s_0 = 4 m_p^2 \approx \mathrm{3.521\ GeV}^2,
\label{scale}
\end{equation}
with $m_p$ the proton mass (see \cite{ms13} for discussions on this choice).

From Eqs. (\ref{st}-\ref{rho}), the analytic results for $\rho(s)$ have been obtained by means of 
even and odd singly subtracted dispersion relations
(integral or derivative forms \cite{am04}):

\begin{eqnarray}
\rho(s) &=& \frac{1}{\sigma_{\mathrm{tot}}(s)}
\left\{  
- a_1\,\tan \left( \frac{\pi\, b_1}{2}\right) \left[\frac{s}{s_0}\right]^{-b_1} \right. \nonumber \\
&+& \left.
\tau \, a_2\, \cot \left(\frac{\pi\, b_2}{2}\right) \left[\frac{s}{s_0}\right]^{-b_2} \right. \nonumber \\
&+& \left. \frac{\pi A}{2} + \pi B \ln\left(\frac{s}{s_0}\right) \right\}.
\label{rhomodel}
\end{eqnarray}

We note that these parameterizations, denoted RRL1L2, are analytically
similar to the COMPETE model RRPL2, where P stands for a critical
Pomeron (constant Pomeranchuck component) \cite{compete1}. The differences
concern: a) the phenomenological interpretation of the singularities
(single or double poles), which may not be relevant; b) the presence
of the free parameter $A$ also in the $\rho(s)$ result; c)
the energy scale $s_0$, which is a free fit parameter in the COMPETE
analysis. We also recall that the logarithmic terms, L1 and L2, are present 
in the Block and Halzen parametrization (fixed energy scale) \cite{bh}.

The RRL1L2 model has only 6 free fit parameters,
$a_1$, $b_1$, $a_2$, $b_2$,  
$A$, and $B$, which are determined through fits to the experimental data 
on $\sigma_{tot}$ and $\rho$ data
from $pp$ and $\bar{p}p$
elastic scattering in the interval 5 GeV - 13 TeV.

\section{Fits and Results}

\subsection{Ensembles and Data Reductions}

The data below 7 TeV have been collected from the PDG database \cite{pdg16}, without any
kind of data selection or sieve procedure (we have used all the
published data by the experimental collaborations). The data at 7 and 8 TeV
by the TOTEM and ATLAS Collaborations can be found in Table 1 in \cite{fms17b},
together with further information and complete list of references;
the data at 13 TeV, Eq. (\ref{totem13}), are also included.

Given the tension between the TOTEM and ATLAS measurements on $\sigma_{tot}$
at 7 TeV and mainly 8 TeV, we shall consider three ensembles of
$pp$ and $\bar{p}p$ data above 5 GeV, both comprising the same dataset
in the region below 7 TeV. We then construct:

\vspace{0.2cm}

\noindent
Ensemble TOTEM (denoted T) by adding only the TOTEM data in the interval 7 - 13 TeV;

\vspace{0.2cm}

\noindent
Ensemble ATLAS (A) by adding only ATLAS data at 7 and 8 TeV;

\vspace{0.2cm}

\noindent
Ensemble TOTEM + ATLAS (T + A)  by adding all the TOTEM and ATLAS data at 7, 8 and 13 TeV.

\vspace{0.2cm}

The data reductions were performed
with the objects of the class TMinuit of ROOT Framework 
and using the default MINUIT error analysis \cite{minuit}. We have carried out global fits using a $\chi^{2}$ fitting procedure,
where the value of $\chi^{2}_{min}$ is distributed as a $\chi^{2}$ distribution with $\nu$ degrees of freedom. The global fits to $\sigma_{tot}$ and $\rho$ data were performed adopting an interval $\chi^{2}-\chi^{2}_{min}$ corresponding, in the case of normal errors, to the projection of the $\chi^{2}$ hyper-surface containing 90\% of probability; this corresponds to $\chi^{2}-\chi^{2}_{min}=10.65$ (for 6 free parameters). 

As a convergence criteria we consider only minimization result which imply positive-definite covariance matrices, since theoretically the covariance matrix for a physically motivated function must be positive-definite at the minimum. 
As tests of goodness-of-fit we shall consider the chi-square per degree of freedom,
$\chi^2/\nu$, and the integrated probability, $P(\chi^2)$ \cite{bev}.

\subsection{Fit Results}

The  fit results are displayed in Table 1.
Within CL of 90 \%, we have evaluated the uncertainty regions through error
propagation from the fit parameters. The results with ensembles T, A and T + A
are shown in Figures 1, 2 and 3, respectively.

\end{multicols}

\begin{table}[H]
\centering
\caption{Fit results with model RRL1L2, Eqs. (\ref{stmodel}-\ref{rhomodel}). }
\begin{tabular}{c@{\quad}c@{\quad}c@{\quad}c@{\quad}c@{\quad}}
\hline 
Ensemble: & T  & A & T$\,+\,$A  \\
\hline

$a_1$  (mb)  & 58.6$\,\pm\,$1.7  & 59.1$\,\pm\,$1.7      & 58.8$\,\pm\,$ 1.6  \\

$b_1$    & 0.226 $\,\pm\,$ 0.018  & 0.238 $\,\pm\,$ 0.039  & 0.231$\,\pm\,$ 0.017 \\
 
$a_2$  (mb)  & 17.0$\,\pm\,$2.3  & 17.1$\,\pm\,$2.3      & 17.1$\,\pm\,$2.3   \\

$b_2$     & 0.547$\,\pm\,$0.039   & 0.549$\,\pm\,$0.040    & 0.548$\,\pm\,$0.039 \\

$A$   (mb)   & 3.62$\,\pm\,$0.37  & 3.97$\,\pm\,$0.35      & 3.76$\,\pm\,$0.33    \\

$B$ (mb)   & 0.135$\,\pm\,$0.027  & 0.101$\,\pm\,$0.026    & 0.122$\,\pm\,$0.022 \\

\hline
$\nu$            &   249            &   236           &     251     \\
$\chi^{2}/\nu$   & 1.210            &  1.136          & 1.238 \\
$P(\chi^2)$       & 1.2 $\times$ 10$^{-2}$  & 7.4 $\times$ 10$^{-2}$ & 6.1 $\times$ 10$^{-3}$  \\
\hline 
\end{tabular}
\label{t2}
\end{table}


\begin{figure}[H]
\begin{center}
 \includegraphics[width=8.0cm,height=7cm]{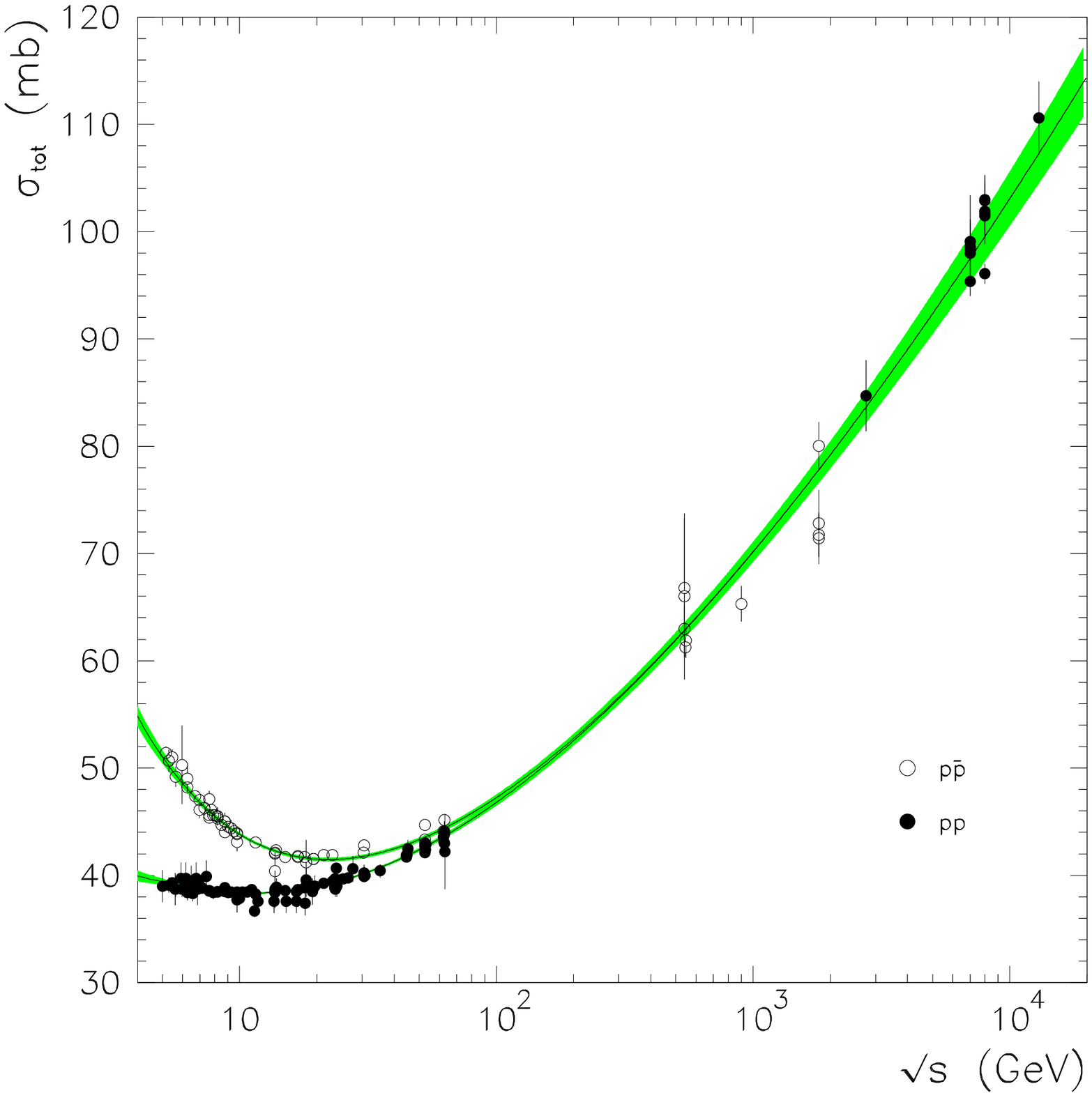}
 \includegraphics[width=8.0cm,height=7cm]{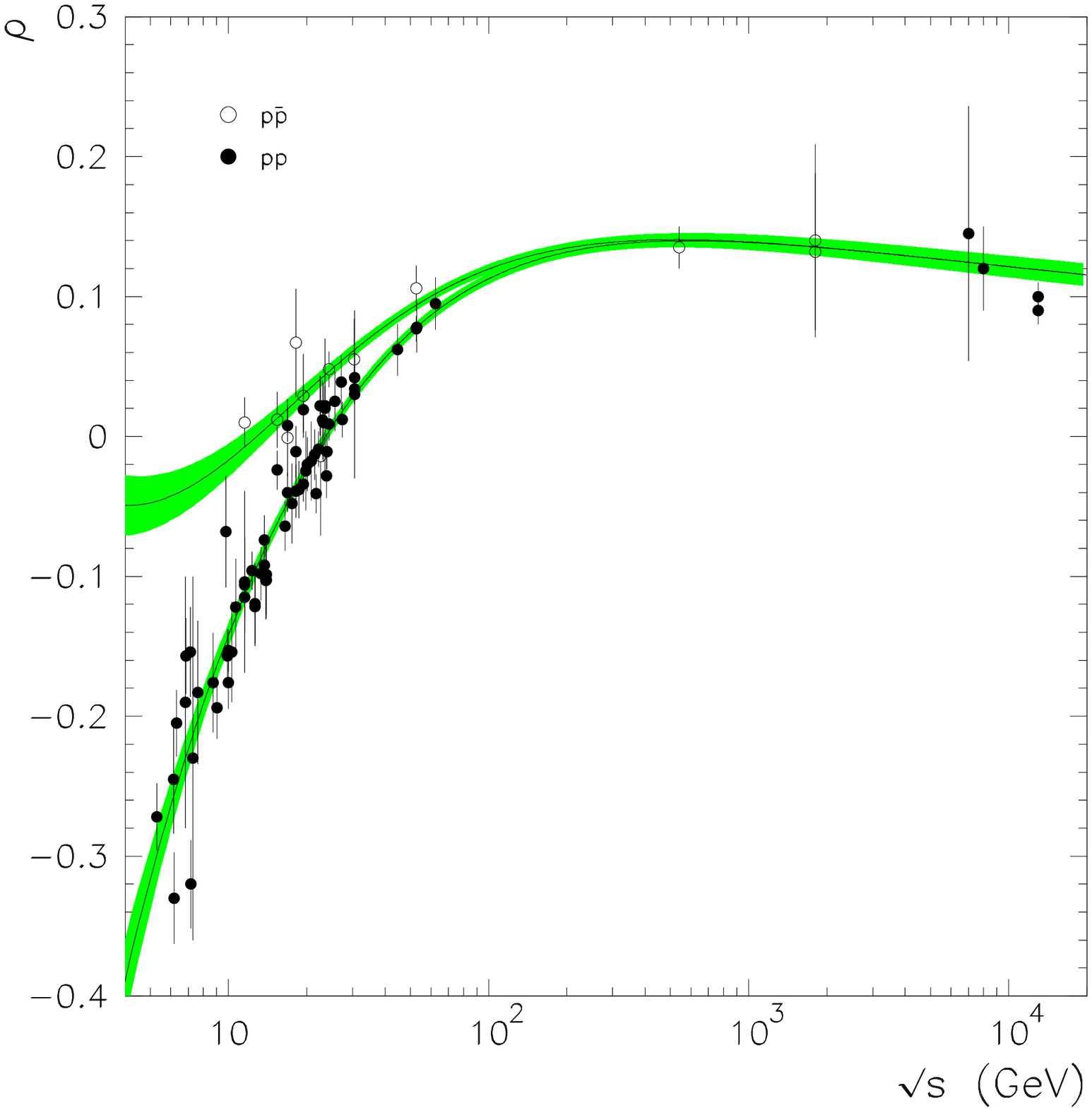}
 \caption{Fit results with Model RRL1L2 (Table 1) to $\sigma_{tot}$ and $\rho$
data from ensemble T.}
\label{f1}
\end{center}
\end{figure}

\begin{figure}[H]
\begin{center}
 \includegraphics[width=8.0cm,height=7cm]{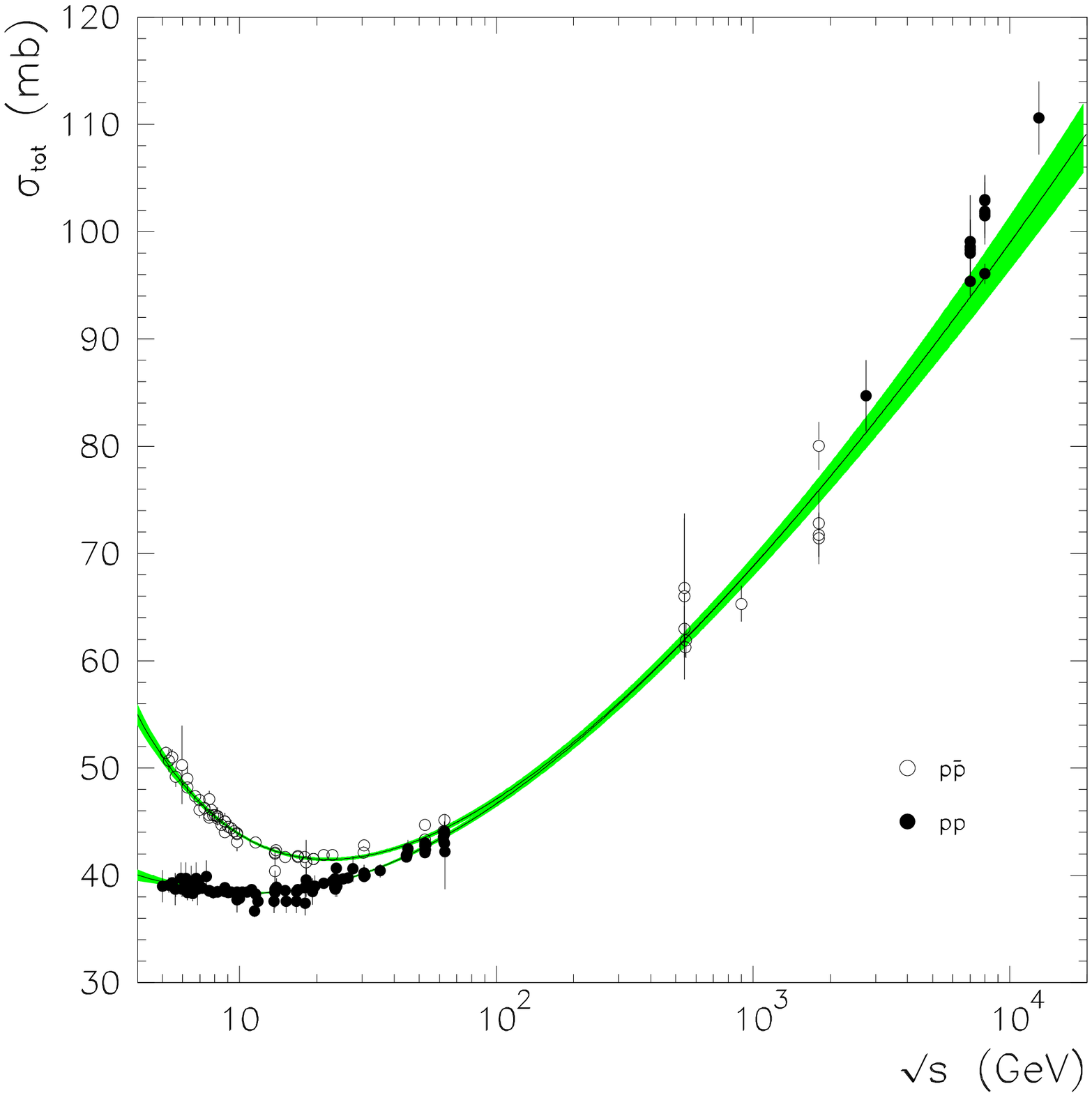}
 \includegraphics[width=8.0cm,height=7cm]{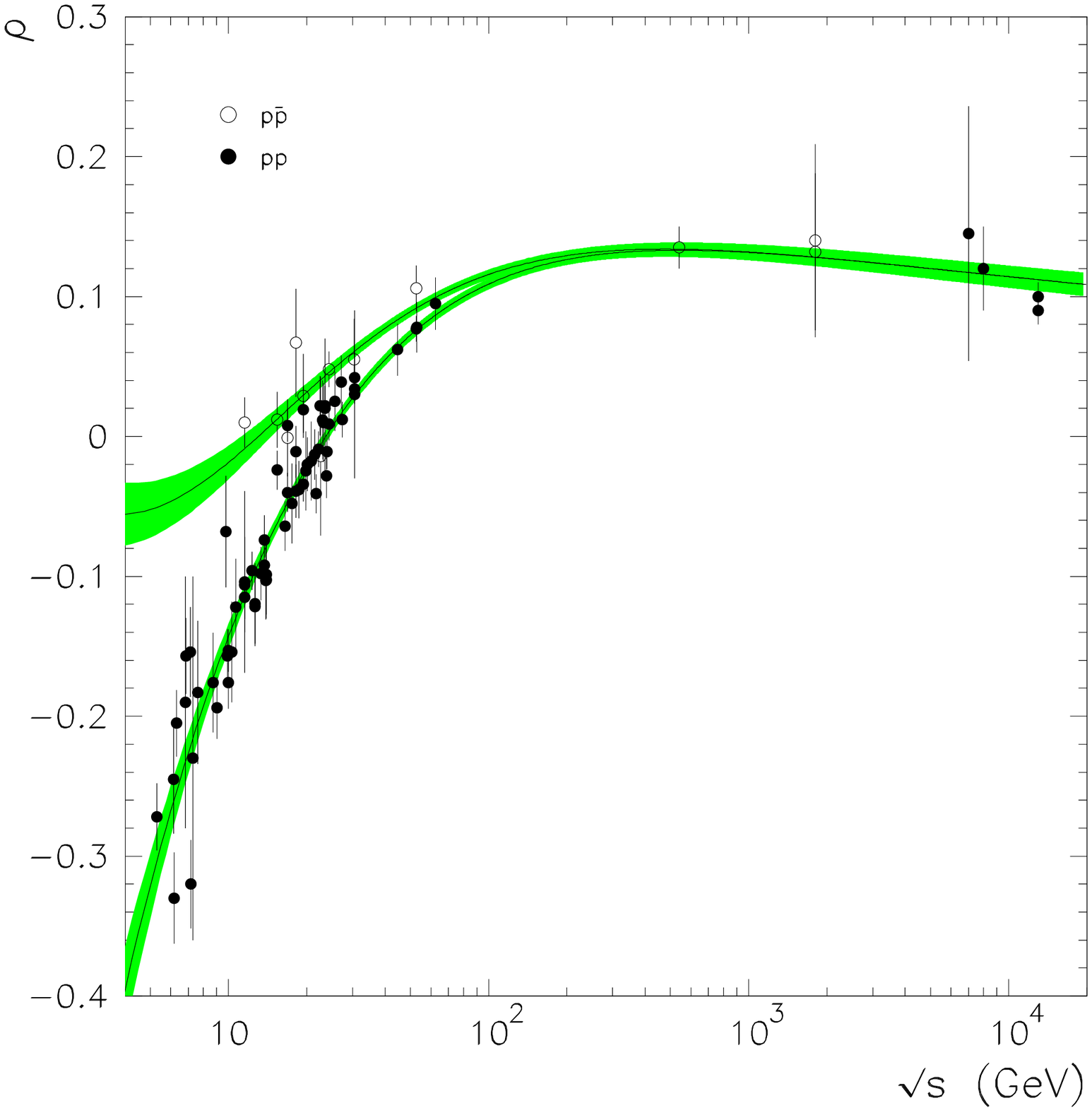}
 \caption{Fit results with Model RRL1L2 (Table 1) to $\sigma_{tot}$ and $\rho$ data
from ensemble A.}
\label{f2}
\end{center}
\end{figure}

\begin{figure}[H]
\begin{center}
 \includegraphics[width=8.0cm,height=7.5cm]{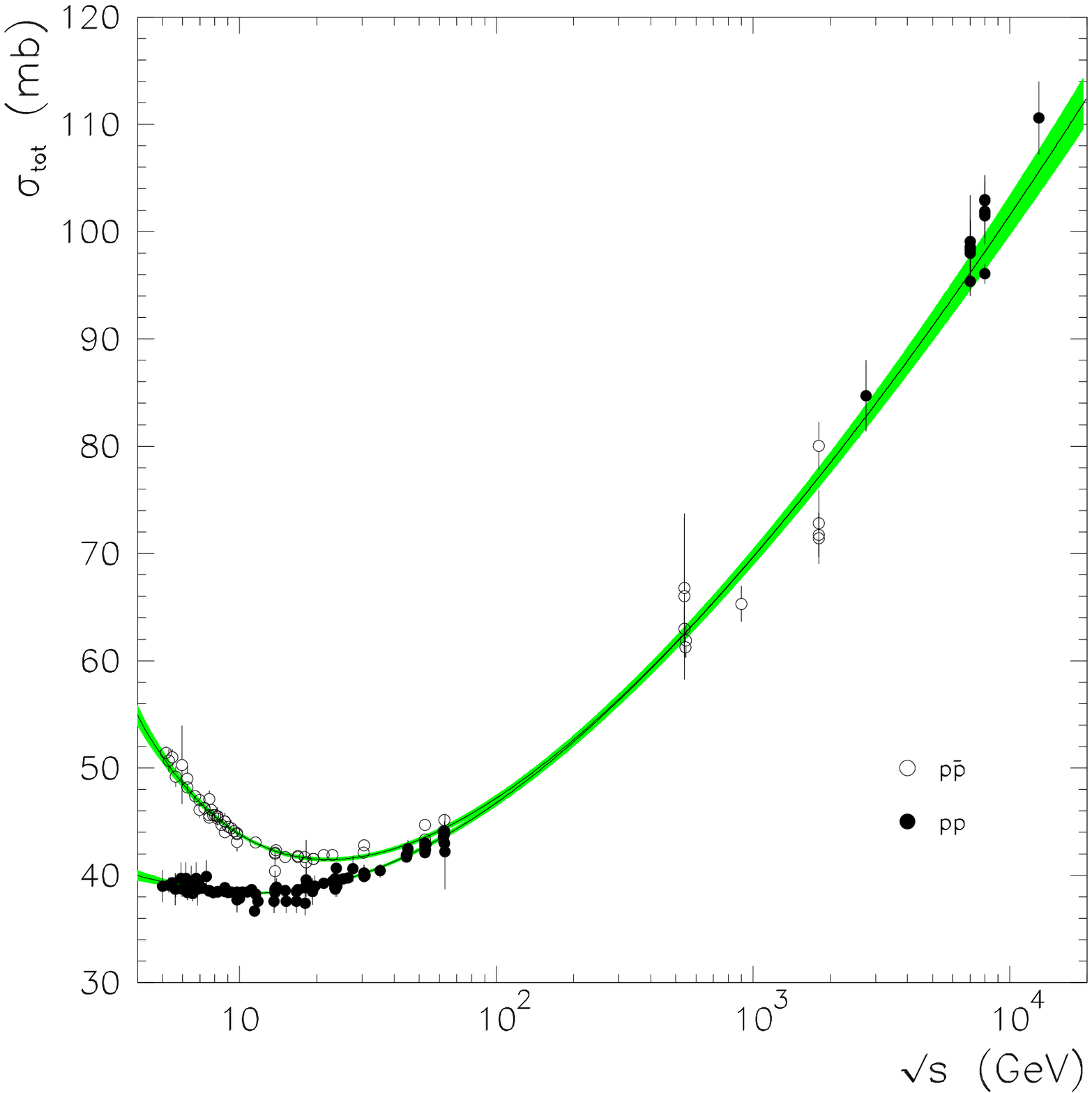}
 \includegraphics[width=8.0cm,height=7.5cm]{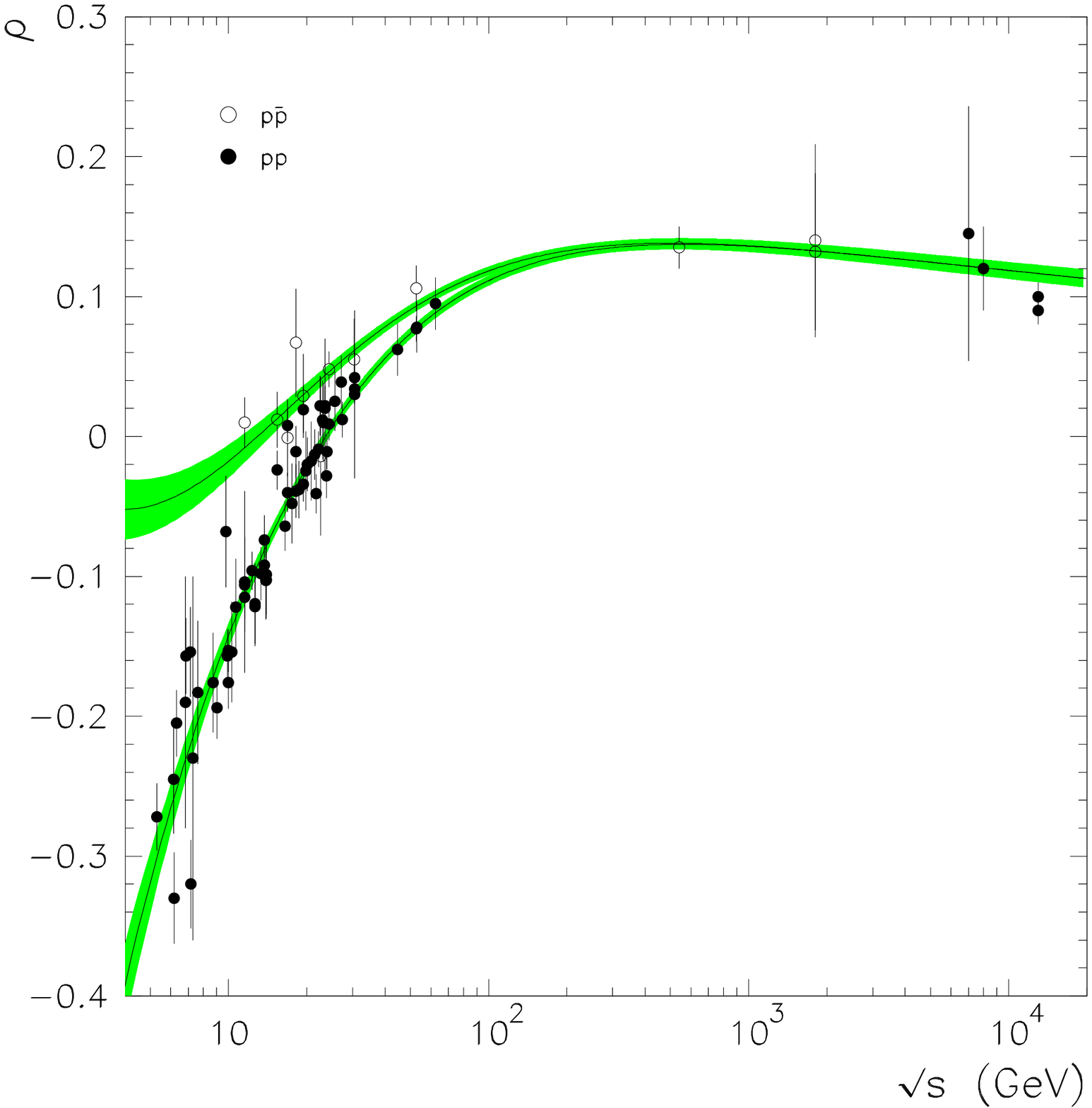}
 \caption{Fit results with Model RRL1L2 (Table 1) to $\sigma_{tot}$  and $\rho$ data
from ensemble T + A.}
\label{f3}
\end{center}
\end{figure} 

\begin{multicols}{2}

\section{Discussion}

Before discussing the results, it is important to note that the three ensembles do not have the same character. 
On the one hand, T and A are
a kind of ``invented"\ ensembles, since they exclude one or another datasets from two 
different experiments. On the other hand, T + A encompasses all the experimental data
presently available, namely all the information provided by the experimentalists from the LHC. 
For this reason let us discuss separately the results obtained with
T and A, followed by those obtained with T + A.

\textit{Ensemble T and Ensemble A.}
It is well known that the TOTEM data indicate a rise of the total cross section
faster than those indicated by the ATLAS Collaboration in the region 7-8 TeV \cite{fms17a}. This effect is clearly
illustrated by the fit results in Figures 1 and 2 with ensembles T and A,
respectively: the $\sigma_{tot}$ datum at 13 TeV is described within ensemble T, but
not within ensemble A.
In each case, the model predictions at 13 TeV read:
$\sigma_{tot} = 107.3 \pm 2.7$ mb and $\rho = 0.1191 \pm 0.0078$ (ensemble T)
and $\sigma_{tot} = 102.8 \pm 2.7$ mb and $\rho = 0.1121 \pm 0.0081$ (ensemble A).

\textit{Ensemble T + A.}
Taking into account all the experimental data presently available, the fit results with
ensemble T + A is presented in Figure 3. We note that at 13 TeV, the upper uncertainty 
region reaches the lower
error bar of the $\sigma_{tot}$ datum and the
lower uncertainty region barely reaches the upper error bar of the of $\rho$ data.
In fact, at 13 TeV the model predictions read
\begin{eqnarray}
\sigma_{tot} &=& 105.6 \pm 2.1\ \mathrm{mb}, \nonumber \\
\rho &=& 0.1164 \pm 0.0061.
\label{ta}
\end{eqnarray}

All the aforementioned predictions at 13 TeV with ensembles T, A and T+A are
schematically displayed in Fig. 4, together with the TOTEM data.

Yet, in case of ensemble T+A, from Table 1, although the goodness of the fit is not good,
$\chi^{2}/\nu$ = 1.238, $P(\chi^2)$ = 6.1 x 10$^{-3}$ for $\nu$ = 251,
we notice that we did not use any kind of data selection and moreover
the ensemble includes, for the first time, the ATLAS data at 7 and 8 TeV and the
TOTEM data at 13 TeV.

Based on the above discussion, we understand that the fit result with
ensemble T+A suggests that model RRL1L2
may not be excluded by the bulk of experimental data presently available.
Further arguments in this direction are presented in what follows.

\begin{figure}[H]
\begin{center}
 \includegraphics[width=8.0cm,height=9.0cm]{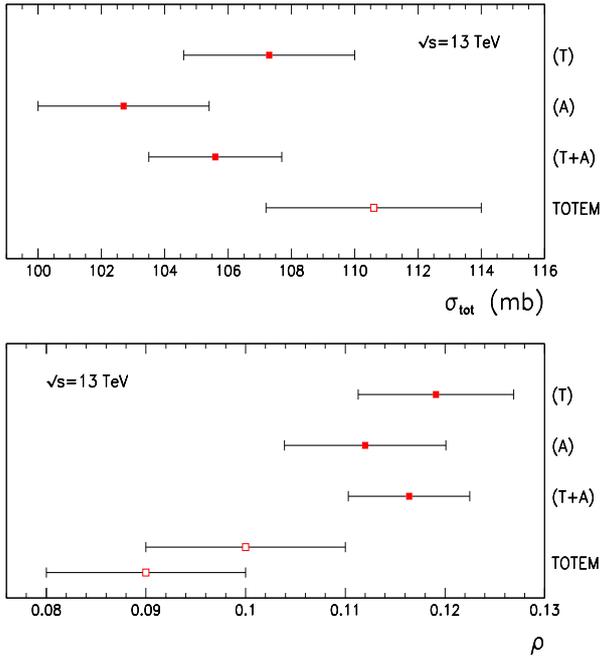}
 \caption{Model predictions for $\sigma_{tot}$ and $\rho$ at 13 TeV from fits to ensembles T, A and T+A
(filled squares),
together with the TOTEM measurements (\ref{totem13}) (empty squares).}
\label{f4}
\end{center}
\end{figure} 

It is important to stress a central point in our
analysis and on the strategy employed
(see also Appendix A.2 in \cite{fms17b} for further discussions and complete list
of reference to the experimental data to be quoted).
In the recent paper by Martynov and Nicolescu, the authors did not
include the ATLAS data at 7 and 8 TeV, because these points ``are incompatible
with the TOTEM data and their inclusion would obviously compromise the
coherence of the overall data"\ \cite{martynico}. The argument is based
on the fact that ATLAS provided only one point at 7 TeV and one point at 8 TeV for
the total cross section, which contrasts with the 4 points at 7 TeV and 5 points
at 8 TeV by TOTEM, all consistent among them at each energy. The incompatibility
can be exemplified by comparison of  the ATLAS result at 8 TeV \cite{atlas8} and the
latest TOTEM measurement at this energy \cite{totem16}, which differ by:
\begin{eqnarray} 
\frac{\sigma_{\mathrm{TOTEM}} - \sigma_{\mathrm{ATLAS}}}{\Delta \sigma_{\mathrm{TOTEM}}} 
= \frac{103.0 - 96.07}{2.3}
= 3.
\label{ta8}
\end{eqnarray} 
Certainly, there may be some missing systematic effect involved , which is expected to be
identified through further analyses.

However, it is important to recall that the situation is not so different from the inconsistencies characterizing the
experimental information at the highest energy reached in $\bar{p}p$ scattering. Indeed, at 1.8 TeV, 
the differences between the CDF Collaboration and the E710 and E811 Collaborations can be
estimated as 2.3 standard deviation:
\begin{eqnarray} 
\frac{\sigma_{\mathrm{CDF}} - \sigma_{\mathrm{E710}}}{\Delta \sigma_{\mathrm{E710}}} = 2.3,
\label{tevatron}
\end{eqnarray} 
suggesting also some missing systematic uncertainty effect which, however, was never identified.

As a consequence, except for some particular studies excluding one or another set \cite{alm03,lm03,lmm04},
most analyses consider the complete dataset with the three points at 1.8 TeV. As a further curious consequence, most analyses are
not compatible with none of them, since the curves lie between the CDF datum (upper) and
the E710/E811 data (lower). This is a characteristic behavior present in the majority of
phenomenological approaches and also in the COMPETE, PDG,
Martynov-Nicolescu analyses and obviously in our own work (Figures 1, 2 and 3).
 
As already mentioned, in case of the LHC data, it is expected that the discrepancies might be resolved
through further analyses and new data. However,
we would like to call the attention to the possibility that the systematic differences
between TOTEM and ATLAS could remain, even after further and detailed re-analyses.
In this case, it would be difficult to carry out forward amplitude analyses (data reductions
through analytic parameterizations for $\sigma_{tot}$ and $\rho$), without taking account of the 
bulk of experimental information available from the LHC, namely all the TOTEM and ATLAS results.

Within this possible scenario, the analysis and results here presented
may have an important role for future investigation, since they suggest that
the Pomeron dominance may be not excluded by the experimental data presently available.
In fact, by comparing the predictions of the RRL1L2 model within ensemble T+A at 13 TeV, (\ref{ta}), with the TOTEM 
measurements, (\ref{totem13}),
we obtain:
\begin{eqnarray} 
\frac{\sigma_{\mathrm{TOTEM}} - \sigma_{\mathrm{predic}}}{\Delta \sigma_{\mathrm{TOTEM}}} = 
\frac{110.6 - 105.6}{3.4} = 1.47,
\nonumber
\end{eqnarray} 
\begin{eqnarray} 
\frac{\rho_{\mathrm{predic}} - \rho_{\mathrm{TOTEM}}}{\Delta \rho_{\mathrm{TOTEM}}} = 
\frac{0.1164 - 0.095}{0.010} = 2.14.
\nonumber
\end{eqnarray} 
Therefore, the differences are smaller than those associated with the TOTEM - ATLAS at 8 TeV,
 (\ref{ta8}),
and with the CDF - E710/E811 difference at 1.8 TeV, (\ref{tevatron}).
We understand that, in the experimental context presently available, these facts
corroborate the effectiveness of the model, the importance of the ensembles and 
the adequacy of the data reduction.

\section{Conclusions and Final Remarks}

We have presented a forward amplitude analysis on the experimental data 
from $pp$ and $\bar{p}p$ scattering in the energy region
from 5 GeV up to 13 TeV. 
We have used  analytic parameterizations for $\sigma_{tot}(s)$
and $\rho(s)$ characterized by Pomeron dominance at the highest energies,
represented by double and triple poles.
Up to our knowledge, this is the first quantitative analysis including
in the data reductions all the experimental data presently available. 

Based on the fit results and taking into account both, theoretical and experimental
uncertainties, we have argued that the RRL1L2 model may not be excluded by the
bulk of experimental data.

We notice that this RRL1L2 parametrization, may not be the
best representative approach for a Pomeron model in forward scattering.
The main point was to show that even a simple parametrization, with only 6 free fit parameter
and even (under crossing) leading contributions, may not be excluded in fits to a dataset including all the
experimental information that have been obtained at the LHC on $\sigma_{tot}$ and $\rho$.

We are presently investigating different forward Pomeron models\footnote{Some
preliminary results have already been presented in \cite{blm18a}.},
now taking into account: (1) confidence levels with one and two standard deviations
(68.6 \% and 95.5 \%, respectively); (2) the introduction of one more free parameter,
represented by the subtraction constant in singly subtracted dispersion relations. 
The analysis is in progress and the results shall be reported elsewhere \cite{blm18c}.

Certainly, to understand and/or to resolve the tension between the TOTEM and ATLAS
data is a crucial point for amplitude analyses and unquestionable conclusions.
In this direction, beyond further re-analysis, measurements on both $\sigma_{tot}$ and $\rho$
at 13 TeV by the ATLAS Collaboration, may bring new insights on the subject.
In conclusion,
at this stage, it may still be premature to 
exclude one or another set of data from different experiments.


\section*{Acknowledgments}

This research was partially supported by the Conselho Nacional de Desenvolvimento Cient\'{\i}fico e Tecnol\'ogico (CNPq) and by the Funda\c{c}\~ao de Amparo \`a Pesquisa do Estado do Rio Grande do Sul (FAPERGS). EGSL acknowledge the financial support from the Rede Nacional de Altas Energias (RENAFAE).


\end{multicols}


\begin{thebibliography}{h!}

\bibitem{pred} 
V. Barone and E. Predazzi, \textit{High-Energy Particle Diffraction},
Spring-Verlag, Berlin (2002).


\bibitem{collins}
P.D.B. Collins, \textit{An Introduction to Regge Theory \& High Energy Physics},
Cambridge University Press, Cambridge (1977).


\bibitem{fr}
J. R. Forshaw, D. A. Ross, \textit{Quantum Chromodynamics and the Pomeron},
Cambridge University Press, Cambridge (1997).

\bibitem{land}
S. Donnachie, G. Dosch, P.V. Landshoff and O. Natchmann,
{\it Pomeron Physics and
QCD}, Cambridge University Press, Cambridge (2002).


\bibitem{fms17b}
D.A. Fagundes, M.J. Menon, P.V.R.G. Silva,
Int. J. Mod. Phys. A 32 (2017) 1750184.


\bibitem{dl13}
A. Donnachie and P.V. Landshoff, 
Phys. Lett. B 727 (2013) 500.

\bibitem{compete1}
COMPETE Collaboration, J. R. Cudell et al., 
Phys. Rev. D 65, (2002) 074024.

\bibitem{compete2} 
COMPETE Collaboration, J. R. Cudell et al.,  
Phys. Rev. Lett. 89, N. 20, (2002) 201801.

\bibitem{pdg16}
Particle Data Group, C. Patrignani et al., 
Review of Particle Physics,
Chin. Phys. C 40 (2016) 100001.



\bibitem{totem1}
TOTEM Collaboration, G. Antchev et al.,
preprint CERN-EP-2017-321; arxiv:1712.06153

\bibitem{totem2}
TOTEM Collaboration, G. Antchev et al.,
preprint CERN-EP-2017-335.

\bibitem{odderon}
L. Lukazsuk, B. Nicolescu, 
Lett. Nuovo Cim. 8 (1973) 405.

\bibitem{oddname}
D. Joynson, E. Leader, B. Nicolescu, C. Lopez,
Nuovo Cim. A 30 (1975) 345.

\bibitem{agn} 
R. Avila, P. Gauron, B. Nicolescu,
Eur. Phys. J. C 49 (2007) 581.



\bibitem{martynico}
E. Martynov, B. Nicolescu, Phys. Lett. B 778 (2018) 414;
arXiv:1711.03288 [hep-ph].

\bibitem{atlas7}
ATLAS Collaboration, G. Aad  et al., Nucl. Phys. B 889 (2014) 486.

\bibitem{atlas8}
ATLAS Collaboration, M. Aaboud  et al., Phys. Lett. B 761 (2016) 158. 

\bibitem{fms17a}
D.A. Fagundes, M.J. Menon, P.V.R.G. Silva, 
Nucl. Phys. A 966 (2017) 185. 


\bibitem{kmr1}
V.A. Khoze, A.D. Martin, M.G. Ryskin, Phys. Rev. D 97 (2018) 034019.


\bibitem{kmr2}
V.A. Khoze, A.D. Martin, M.G. Ryskin, Phys. Lett. B 780 (2018) 352.


\bibitem{ms13}
M.J. Menon, P.V.R.G. Silva, 
J. Phys. G 40 (2013) 125001; Erratum: J. Phys. G 41 (2014) 019501.


\bibitem{am04}
R. F. \'Avila and M. J. Menon,  
Nucl. Phys. A 744 (2004) 249.

\bibitem{bh}
M. M. Block and F. Halzen, 
Phys. Rev. D 86 (2012) 051504.


\bibitem{minuit}
F. James, MINUIT Function Minimization and Error Analysis, Reference Manual,
Version 94.1, CERN Program Library Long Writeup D506 (CERN, Geneva, Switzerland, 1998).


\bibitem{bev}
P. R. Bevington and D. K. Robinson, Data Reduction and Error Analysis
for the Physical Sciences (Boston, MA: McGraw-Hill, 1992).

\bibitem{totem16}
TOTEM Collaboration, G. Antchev et al., 
Eur. Phys. J. C 76 (2016) 661.



\bibitem{alm03}
R.F. \'Avila, E.G.S. Luna, M.J. Menon,
Phys. Rev. D 67 (2003) 054020.


\bibitem{lm03}
E.G.S. Luna, M.J. Menon,
Phys. Lett. B 565 (2003) 123.

\bibitem{lmm04} 
E.G.S. Luna, M.J. Menon, J. Montanha,
Nucl. Phys. A 745 (2004) 104.


\bibitem{blm18a}
M. Broilo, E.G.S. Luna, M.J. Menon, Leading Pomeron Contributions and the TOTEM Data at 13 TeV,
proceedings XIV Hadron Physics - 2018, arXiv:1803.06560 [hep-ph].


\bibitem{blm18c}
M. Broilo, E.G.S. Luna, M.J. Menon, Forward Elastic Scattering and Pomeron Models
(in preparation).



\end{thebibliography}
\end{document}